\documentstyle[12pt]{article}
\begin{document}

\begin{titlepage}

\title{The Interrelation of a Z(3) Gauge Theory on the\\[2mm]
        Flat Lattices and a Spin-1 BEG Model}

\author{N.S.~Ananikian\thanks{E-mail address: nanan@tpd.erphy.armenia.su}
\\[2mm]
{\small \sl Department of Theoretical Physics,
            Yerevan Physics Institute,} \\
{\small \sl Alikhanian Br. 2, 375036 Yerevan, Armenia} \\[2mm]
and \\[2mm]
R.R.~Shcherbakov\thanks{E-mail address: shcher@thsun1.jinr.dubna.su}
                \thanks{On leave of absence from
               Department of Theoretical Physics,
               Yerevan Physics Institute, Armenia} \\[2mm]
{\small \sl Bogoliubov Laboratory of Theoretical Physics,} \\
{\small \sl JINR, 141980 Dubna, Russia}
}

\date{\nonumber}
\maketitle

\bigskip
\begin{abstract}
The Z(3) gauge model with double plaquette representation
of the action on the flat triangular and square lattices is constructed.
It is reduced to the spin-1 Blume-Emery-Griffiths (BEG) model.
An Ising-type critical line of a second-order phase transition
is found.
\end{abstract}
\bigskip

\thispagestyle{empty}
\end{titlepage}

\newpage
\setcounter{page}2
\normalsize
\section{Introduction}

It is well known that symmetries play a key role in modern
physics and have a direct relation with universality
hypothesis. The power of universality ideas comes from the fact
that they relate the critical properties of a gauge theory
in ($d+1$) space-time dimensions to those of a classical scalar
field theory (or spin system) in $d$ spatial dimensions and
allow predictions of critical indices in theories with
continuous phase transitions.

The lattice gauge theory with one-parameter representation of the
action was introduced by Wilson~\cite{Wilson}:
\begin{equation}
\label{Wilson}
S=-\beta \sum_p \Re eU_p\,,
\end{equation}
where $U_p$ denotes the usual plaquette variable, the product of
link gauge fields $U_{x,\mu}$ around a plaquette. The critical
behavior of this model depends on the dimensionality and the
gauge group and includes both first- and second-order phase
transitions.

The enlarged lattice gauge actions involving new double plaquette
interaction terms
were proposed and studied in $3d$ and $4d$ by
Edgar~\cite{Edgar}, Bhanot~et al~\cite{Bhanot}. The $2d$
version of one of these lattice gauge models with
Z(2) gauge symmetry formulated on the planar rectangular
windows was investigated by Turban~\cite{Turban}.
This model with pure gauge action had been reduced to the
usual spin-$\frac{1}{2}$ Ising model on the square lattice
and the point of a second-order phase transition was found.

In our  paper we constructed the Z(3) gauge model on the flat
triangular and square lattices with double plaquette (window)
representation of the action. The choice of this mixed action
allowed us to connect it with the Hamiltonian of  the  spin-1
BEG model~\cite{BEG} and receive the line of the second-order
phase transition.

\section{The BEG model}

The BEG model plays an important role in the development of the
theory of multicritical points. Its Hamiltonian is
\begin{equation}
\label{BEG}
-\beta H = \sum_{<i,\,j>}\left[
J\,S_iS_j\,+\,K\,S_i^2S_j^2 \right]\,-\,
\Delta\sum_iS_i^2\,,
\end{equation}
where $S_i=0,\pm 1$ is the spin variable at the site $i$ and
$<i,j>$ designate the nearest-neighbor pairs of the lattice
sites. This model exhibits a rich critical behavior including
the first- as well as second-order phase transitions and
reflects interesting phenomena connected with real physical
systems such as the phase separation in ${}^3{\rm He}-{}^4{\rm He}$
mixture, multicomponent fluids, microemulsions and so on.
Its phase diagrams have been investigated by means of
mean-field~\cite{Hoston}, renormalization group~\cite{Berker} and other
approximations. The exact results have been obtained on the
honeycomb~\cite{Hor,Wu} and on the Bethe~\cite{Avakian,Shcher}
lattices.

As to the exact result on the honeycomb lattice, it is limited
to the certain subspace of the exchange-interaction constants
\begin{equation}
\label{Horig}
e^K \cosh\, J = 1\,.
\end{equation}
This condition permits to map the BEG model to the usual
spin-$\frac{1}{2}$ Ising model and get the $\lambda$-line
of the second-order phase transition.

In our previous paper~\cite{Shcher} we solved the BEG model
with condition~(\ref{Horig}) on the Bethe lattice and found
the tricritical point of the second-order phase  transition
for the lattices with coordination number greater than 6.

Now it is interesting to apply the results obtained for the
BEG model to the gauge theory, because of the deep interrelation
between spin and gauge lattice models.

\section{The Z(3) gauge model}

The model is considered
on the flat triangular and square lattices in terms
of the bond variables $U_b$ which take their values in the Z(3),
the group of the third roots of unity.
Let $U_{p_i}=\prod_{b\in\partial p}U_b$ denote the product of
$U_b$'s around an elementary plaquette $i$.

The action of the model is
\begin{equation}
\label{sg}
S_{Gauge}(\beta_{2g},\beta_{2g}',\beta_{g}) =
            S_{p\,p} + S_{p}\,,
\end{equation}
where
\begin{displaymath}
S_{p\,p} = -\sum_{<p_i\,p_j>}
\left\{\beta_{2g} \left( \delta_{U_{p_i},1}\delta_{U_{p_j},1}+
                         \delta_{U_{p_i},z}\delta_{U_{p_j},z} \right)+
 \beta_{2g}' \left( \delta_{U_{p_i},1}\delta_{U_{p_j},z}+
		         \delta_{U_{p_i},z}\delta_{U_{p_j},1} \right)
\right\}\,,
\end{displaymath}
\begin{displaymath}
S_{p} = \beta_{g}\sum_{p_i}
\left( \delta_{U_{p_i},1}+\delta_{U_{p_i},z}\right)\,.
\end{displaymath}
The first summation is over all nearest-neighbor plaquettes and
the second one is over all plaquettes of the lattice,
$z=\exp(i\frac{2\pi}{3})\in{\rm Z(3)}$.

Introducing spin variables $S_i$ in the sites of the dual lattice
such that
\begin{equation}
\label{ss}
\begin{array}{c}
S_i  =\delta_{U_{p_i},1}-\delta_{U_{p_i},z}\,,\\
S_i^2=\delta_{U_{p_i},1}+\delta_{U_{p_i},z}
\end{array}
\end{equation}
the action~(\ref{sg}) becomes
\begin{equation}
\label{sspin}
S_{Spin}(\beta_{2g},\beta_{2g}',\beta_{g})=
-\sum_{<ij>}\left\{
\frac{\beta_{2g}-\beta_{2g}'}{2} S_i S_j +
\frac{\beta_{2g}+\beta_{2g}'}{2} S_i^2 S_j^2
\right\} +
\beta_g\sum_i S_i^2
\end{equation}
in which we recognize the Hamiltonian multiplied by $1/k_BT$
of the BEG model~(\ref{BEG}), where $J=\frac{1}{2}(\beta_{2g}-\beta_{2g}')$,
$K=\frac{1}{2}(\beta_{2g}+\beta_{2g}')$ and $\Delta=\beta_g$.

The corresponding partition function of the model~(\ref{sg}) on the
triangular lattice is
\begin{equation}
\label{zgauge}
Z_{Gauge}^{Triangular}(\beta_{2g},\beta_{2g}',\beta_{g})=
\sum_{\{U\}}\exp\left[
-S_{Gauge}(\beta_{2g},\beta_{2g}',\beta_{g})\right]
\end{equation}
where the sum is taken over all possible configurations of
the gauge variables $\{U\}$. This partition function can be rewritten
in terms of the spin variables $S_i$ defined in the sites of the dual
lattice (honeycomb)
\begin{equation}
\label{zgt}
Z_{Gauge}^{Triangular} = 3^{N/2}Z_{Spin}^{Honeycomb}
\end{equation}
where
\begin{displaymath}
Z_{Spin}^{Honeycomb}=
\sum_{\{S\}}\exp\left[
-S_{Spin} \right]
\end{displaymath}
A factor $3^{N/2}$ has been included in the equation~(\ref{zgt}) to
take into account the difference between the number of gauge $\{U\}$
and spin $\{S\}$ configurations, since for each spin configuration
with N sites we have $3^{N/2}$ identical gauge ones.

The condition~(\ref{Horig}) in terms of the gauge coupling
constants takes the following form
\begin{equation}
\label{relation}
\exp(\beta_{2g}) + \exp(\beta_{2g}') = 2\,.
\end{equation}
Therefore, we can apply the exact solution obtained by Horiguchi and
Wu~\cite{Hor,Wu} for our gauge model and obtain the line of the
second-order phase transition
\begin{equation}
\label{sol1}
\exp(\beta_g)=2\left[
\sqrt{3}\exp(\beta_{2g})-(\sqrt{3}+1)\right]\,.
\end{equation}

In the same way as for triangular lattice we can reduce the partition
function of our model defined on the square lattice to the
partition function of the BEG model on the dual lattice:
\begin{equation}
\label{zgs}
Z_{Gauge}^{Square}(\beta_{2g},\beta_{2g}',\beta_{g})
= 3^{N}Z_{Spin}^{Square}(\beta_{2g},\beta_{2g}',\beta_{g})\,,
\end{equation}
where the factor $3^N$ takes again into account the difference
between the number of spin and gauge configurations.

For the BEG model formulated on the square lattice with
condition~(\ref{relation}) an approximate result have been
obtained by Tang~\cite{Tang} using a free-fermion approximation
technique~\cite{Fan}. Applying this solution for our
model we get the critical line of the second-order phase
transition:
\begin{equation}
\label{sol2}
\exp(\beta_g)=2(\sqrt{2}+1)\left[
\exp(\beta_{2g})-\sqrt{2}\right]\,.
\end{equation}

The critical lines~(\ref{sol1}) and~(\ref{sol2}) of the second-order
phase transition separate the $(\beta_{2g}\,,\beta_g)$ plane onto two
regions as it is shown in the fig.1. We have the phase transition from
the ordered (Z(2) broken) phase which corresponds to confinement to
the disordered one, since confinement in the gauge theory corresponds
to ordering in the spin system.
These transitions belong to the Ising type
and whence they have the same critical exponents.

\section{Conclusion}

We constructed the Z(3) gauge lattice model with double
plaquette (window) representation of the action and showed
that this model is dual to the spin-1 BEG one. Using the
exact solution of the BEG model on the honeycomb lattice
and the solution obtained by the free-fermion approximation
technique on the square lattice we found the lines of the
second-order phase transition.

The constructed action gives possibility to transform  the
Z(3) gauge theory defined on the generalized Bethe lattice
of plaquettes~\cite{Ananikian,Akheyan} to the BEG model on
the  usual  Bethe  lattice  for  which the exact solution was
obtained~\cite{Avakian} and receive exact expressions for the
multicritical points and construct phase diagrams.

\section*{Acknowledgments}

\bigskip


We would like to thank P.~Arnold, R.~Flume, R.~Hairapetyan, K.~Oganessyan,
A.~Sedrakyan and A.~Akheyan for fruitful discussions.

One of us (R.R.S.) wishes to thank V.B.~Priezzhev for the hospitality
extended to him at the N.N.~Bogoliubov Laboratory of Theoretical Physics
of Joint Institute for Nuclear Research.

\medskip
This work was partially supported by German Bundusministerium f\"ur
Furschung und Technologie under grant No. 211-5291 YPI and
by the Grant INTAS-93-633.

\newpage

\ \ \

\vspace{6in}
\begin{center}
Fig. 1.
\end{center}
\medskip
\noindent
The plane of the coupling constants $(\beta_g,\beta_{2g})$ is separated
by the lines of the second-order phase transition onto two regions
(ordered and disordered). The line 1 (2) corresponds to the
triangular (square) lattice.

\end{document}